\documentstyle[twoside,fleqn,aip2col]{article}
\newcommand{\AmS}{{\protect\the\textfont2
  A\kern-.1667em\lower.5ex\hbox{M}\kern-.125emS}}

\def\journal#1&#2(#3){\begingroup \let\journal=\dummyj@urnal
    \unskip, \sl #1\unskip~\bf\ignorespaces #2\rm
    (\afterassignment\j@ur \count255=#3), \endgroup\ignorespaces }
\def\j@ur{\ifnum\count255<100 \advance\count255 by 1900 \fi
          \number\count255 }
\def\dummyj@urnal{%
    \toks@={Reference foul up: nested \journal macros}%
    \errhelp={You forgot & or ( ) after the last \journal}%
    \errmessage{\the\toks@ }}
\def\apsjournal#1&#2,#3(#4){\unskip, #1\ {\bf #2}, #3 (19#4)\unskip}

\def\gsim{\stackrel{>}{\sim}}

\def\*{\hbox{$\rm *$}}

\begin{document}
\hyphenation{author another created financial paper re-commend-ed}

\title{Frontiers of Astrophysics - Workshop summary$^1$
}

%
%

\author{Peter L. Biermann \& Heino Falcke
\address{Max-Planck-Institut f\"ur Radioastronomie\\Auf dem H\"ugel 69\\
D-53121 Bonn\\ Germany}%
}


\begin{abstract}
We summarize recent results presented in the astrophysics session
during a conference on ``Frontiers of Contemporary Physics''. We will
discuss three main fields (High-Energy Astrophysics, Relativistic
Astrophysics, and Cosmology), where Astrophysicists are pushing the
limits of our knowledge of the physics of the universe to new
frontiers.  Since the highlights of early 1997 were the first
detection of a redshift and the optical and X-ray afterglows of
gamma-ray bursts, as well as the first well-documented flares of
TeV-Blazars across a large fraction of the electromagnetic spectrum,
we will concentrate on these topics. Other topics covered are black
holes and relativistic jets, high-energy cosmic rays, $\nu$-Astronomy,
extragalactic magnetic fields, and cosmological models.

\end{abstract}


\maketitle

\section{INTRODUCTION}\footnotetext[1]{Frontiers in Contemporary Physics,
Nashville, May 11-16, 1997, AIP-conference series, Ed. T. Weiler et al.} 
Astrophysics has become one of the most rapidly
evolving disciplines in physics over the last two decades. An aggressive
expansion, mainly driven by new technology, has pushed the limits for the
observer further in photon energy, sensitivity, and spatial
resolution and new sub-disciplines are being added to astronomy and
astrophysics at a breathtaking speed. Young fields like x- and $\gamma$-ray
astronomy already play an integral role in the scientific community,
while TeV- and Neutrino-astronomy are knocking at the door.

Though even the established astronomical sub-disciplines have legions
of their own frontiers, we will here concentrate on the very forefront
of astronomy where astronomy is pushing the very frontiers of physics
itself. Three of those fields were selected for this workshop:

a) High-Energy Astrophysics, where we can study photons and particles
from cosmic accelerators with energies way above what can be produced
in laboratory on Earth---unfortunately, we even do not understand
those cosmic accelerators yet.

b) Relativistic Astrophysics, where we can study General Relativity in
the strong limit never reached in our solar system. Black Holes are
one example, which most astronomers consider already well established
while many physicists wait for the final proof---and there is hope.

c) Cosmology, the most fundamental, yet sometimes also most fantastic
part of physics, where astronomy provides the basic data and sets the
framework for any cosmological model.

Despite the importance of all those questions, a small workshop like
this is of course always dominated by the personal interests of the
scientists present and the and the most recent discoveries. The spring
of 1997 was especially marked by two major new observational
developments:

First, the afterglow of the enigmatic Gamma Ray Bursts (GRBs) was
observed in X-rays, the optical and also in the radio. One also
obtained the first lower limit for the redshift of a GRB,
demonstrating fairly convincingly, that GRBs are cosmological.

Second, Blazars, this fascinating subclass of quasars, were observed
to tens of TeV photon energy (marking the advent of TeV astronomy),
and that with variability of hours. One also observed correlated variability
episodes with the optical and X-rays.

This gives hope that the basic physical mechanisms underlying these
phenomena may be closer to an understanding; they may be even related
after all, and also relate to many other high energy phenomena such as
ZeV cosmic rays and cosmic magnetic fields.

Therefore the authors of this summary decided not to stand in the way
of hot, new trends at a conference like this and gave rather unequal
weight to the different topics, putting here more emphasis on these
new developments.

\section{High-Energy Astrophysics}

\subsection{Gamma-ray bursts (Waxman)}

Gamma Ray Bursts (GRBs), short flashes of gamma-photons on the sky,
have captured the imagination of theorists for decades after they were
initially discovered by satellites to monitor nuclear explosions in
space. Until early 1997 all we had were speculations. This changed
dramatically when the Italian-Dutch satellite BeppoSax started to provide
sufficiently good positions for new GRBs, that for the first time made
radio, optical, and X-ray detections of their afterglow possible.  The
high point of these discoveries was the first redshift of a GRB
\cite{Metzger97,Arav97}, demonstrating  that, in at least one case,
the GRB came from a cosmological distance.

Pre-1997 properties of GRBs have been reviewed in Hartmann
\cite{Hartmann94}, and Fishman \& Meegan \cite{Fishman95}.  They are
isotropically distributed over the sky to all tests made (using the
BATSE data from the Compton Gamma Ray Observatory), with the current
catalogue of over 1100 bursts, have no repeaters as far as can be
stated with certainty, and have a count-rate relation already
suggestive of cosmological distances (a downward deviation from a
--3/2 powerlaw in cumulative numbers at low flux levels).  Their {\it
typical} fluence is $\approx 10^{-5} \, \rm erg/cm^2$, which translates
at cosmological distances to an energy required reminiscent of a
supernova \cite{Piran92}.

The modern theoretical attempts to interpret GRBs center on ideas by
M{\'e}sz{\'a}ros and Rees
\cite{Mesz92,Mesz93,Mesz94,Rees94,Panait97a,Panait97b,Meszaros97a,Meszaros97b},
as well as Paczy{\'n}ski
\cite{Pacz86,Pacz87,Pacz91,Pacz92a,Pacz92b,Pacz93,Pacz94,Pacz97}.  In
these models a relativistic shock is caused by a fireball expanding into 
a surrounding medium, such as the interstellar medium or a stellar
wind (or jet?), accelerating electrons/positrons to a very
high energy, which then produce the electromagnetic radiation observed in GRBs
and their afterglow.  The low level of associated radiation at
other wavelengths limits the baryonic load ({\it i.e.}, the content in
thermal protons, which would be expected to cause thermal radiation)
of the emitting regions to very low amounts, and also constrains the
scale of the emitting region to lengths much larger than a neutron
star.  It appears that all the emission seen is actually non-thermal.

Such models readily lend themselves to the modification to hadronic
particle populations
\cite{Waxman95a,Waxman95b,Vietri95,Vietri96,MU95,MU96,miralda96,Waxman97a,Waxman97b},
so as to accelerate also baryons to very high energy, and use the
ensuing hadronic interactions to produce gammas as as well as high
energy protons.  It then becomes of interest to ask whether a
combination of hypothetical models can be built, which produce both
very high energy cosmic ray particles and the gamma rays which are
observed.

We note that there appears to be no correlation between the arrival
directions of gamma ray bursts \cite{Hartmann96,Stanev96} and the
super-galactic plane ({\it i.e.} a `local' galaxy over-density), in contrast
to the finding for the very high energy cosmic rays
\cite{PRL95,Hayashida96,Uchihori97} (see below). Also, there is no correlation
between the gamma-ray bursts of the third
BATSE catalogue as well as the pre-GRO bursts and the arrival
directions of high
energy cosmic rays from the Haverah Park set, covering
approximately the same period of time \cite{Stanev96}.

Eli Waxman showed in his contribution that now quantitative models for the
emission observed from GRBs are available which can be tested.  He was able to
provide a lengthy discussion of the recently detected GRB 970508, including the
determination of a lower limit to its redshift (see also
\cite{Waxman97a,Waxman97b,Waxman97c,Waxman97d,Waxman97x,Waxman97y}).
Other recent work is described in
\cite{Meszaros97a,Meszaros97b,Panait97a,Panait97b,Stern97a,Stern97b,Vietri97b}
\cite{Arav97,Costa97,Djorgovski97a,Galama97,Guarnieri97,Hurley97}
\cite{Krawczynski97,Mallozzi97,Metzger97,Paradijs97,Piro97}
\cite{Sahu97a,Sahu97b,Vietri97a,Frail97b,Mitra97,Pacz97}
and in IAU circulars.
This list is only exemplary.

We briefly outline the essential features of the fireball model
\cite{Meszaros97a,Waxman97x}, considering for simplicity only the external
shockwave propagating into the surrounding medium:  A large amount of energy 
is released and starts an expansion with a highly relativistic velocity, of 
order $\gamma \approx 300$.  Rather akin to the normal Sedov solution for
supernova remnants, the heated interstellar medium dominates the expansion 
after the initial burst.  There is a shell of relativistic matter with 
radius $r$, and thickness $r/(4 \gamma^2)$ in the observers frame; the 
density in the shell frame is given by $n' = 4 \gamma n$, and the internal 
energy $e' = 4 \gamma^2 n m_p c^2$, disregarding any heavy elements. The 
heated energy of the ISM shell is 
$4 \pi r^2 (r/(4 \gamma^2)) \gamma^2 e' \, = \, E/2$ where $E$ 
is the energy of the explosion, and the factor $1/2$ comes from sharing the
energy between ejecta and the ISM.  The factor $\gamma^2$ in front of the
internal energy in the shock frame $e'$ derives from the transformation to 
the observers frame.  This then gives the evolution of $\gamma$ with time
$\gamma \, \sim \, r^{-3/2}$,
which---again---is rather reminiscent of the Sedov relation for the 
expansion velocity
of a shock in the interstellar medium ({\it e.g.} \cite{Cox72}).
Light is emitted over
$t \, = \, (r/(2 \gamma^2 c) \, \sim \, r^4$, and
the electron population starts at the relativistic thermal energy
$\gamma_{em} \, = \; \xi_e \gamma \frac{m_p}{m_e}$
with $\xi_e$ being some factor of order unity.  The electrons continue 
to higher energies with a powerlaw determined by the shock.  The synchrotron
emission  has its minimum frequency at
$\nu_m \, = \, \gamma \gamma_{em}^2 (e_e B'/(m_e c))$,
where $e_e$ is the charge of the electron.  This translates to
$$
\nu_m \, \sim \, t^{-3/2}
$$
and so the temporal behavior of the spectrum is expected to be
$$
F_{\nu} \, \sim \, (\nu / \nu_m)^{-\alpha} \, \sim \, t^{-3 \alpha/2},
$$
where $\alpha$ is the spectral index of the synchrotron radiation.  This
is a testable case:  For GRB 970228 the spectrum was determined between
optical and X-ray wavelengths and gives $\alpha \approx 0.63$ and consequently
the temporal behavior has a scaling as

$$
F_{\nu} \, \sim \, t^{-0.95}
$$
to be compared with an observed $t^{-1}$ behavior.  This is a magnificent
and positive test and was one of the main points of Eli Waxman's
lecture.  At the time of writing the optical afterglow has been observed
to fade steadily still after six months for GRB970228.

Particles can be accelerated at internal shocks in the GRB outflow, which may
be responsible for the rapid variability in some GRBs \cite{Mesz94}.  Eli
Waxman proposed that in the shocks produced by the explosion both
electrons and protons get  accelerated; the absolute limit to acceleration for
energetic protons  is the spatial constraint, assuming that acceleration is
sufficiently rapid: $
\gamma_p m_p c^2/(e B') \, < \, r/\gamma,
$
where $\gamma$ is the Lorentz factor of the shock and the expression is in
the shock frame.
Scaling the magnetic field with the assumption of equipartition (introducing
the factor $\xi_B$) in this  model we get $
B' \, = \, \xi_B \, \sqrt{8 \pi e'} = \xi_B \, \sqrt{32 \pi \gamma^2 n m_p c^2}.
$
This then leads to a maximum proton energy in the shock frame of

$$
\gamma_{p, max} m_p c^2 \; = \; 2 \cdot 10^7 \, n^{1/2} \, \xi_B \, r_{17} 
\, \rm erg 
$$
using the empirical scaling for the numerical relationship between Lorentz
factor and radius from radio observations \cite{Frail97b}
$
\gamma \; = \; 3 \, r_{17}^{-3/2},
$
where $r_{17} = r/10^{17} \, \rm cm$.  The maximum energy written in the
observers frame is then given by

$$
\gamma_{p,max} m_p c^2\; = \; 3 \cdot 10^{19} \, n^{1/2} \, \xi_B \, 
r_{17}^{-1/2} \, \rm eV
$$
demonstrating that very high energies are possible.  Going to the smaller
radius, where the GRB event first becomes visible, near
$5 \cdot 10^{15} \, \rm cm$, where the Lorentz factor of the shock is near 300,
obviously we have there
$$
\gamma_{p,max,*} \; \approx \; 10^{20} \, n^{1/2} \, \xi_B \, \rm eV
$$
to within the uncertainties sufficient to account for the highest
energies observed in Cosmic rays, if we accept $\xi_B \approx 1$.  Of 
course, everything depends then on which density of the interstellar 
medium is relevant; if we take the density recently determined from ROSAT
observations for the hot interstellar medium
\cite{Snowden97}, of $\approx 3 \cdot 10^{-3} \, \rm cm^{-3}$, then the
maximum energy is too small, but if we take the average density as
determined from neutral hydrogen measurements of order 1 particle per ccm, then
the proposal  becomes viable.  So, Eli Waxman argued  that such high energy
protons may account for the observed ultra high energy cosmic rays.

One serious question, however, is whether the overall energetics of
the fireball are reasonable \cite{Sari97,Dar97} or actually exceed the
level given by any conceivable model of neutron star mergers or other
stellar collapses.

Eli Waxman also went through the exercise to estimate the
contribution to a universal high energy neutrino background, which we
all hope to see soon with AMANDA at the South Pole.

Of course, the latest news from the May 8 burst which had come out the
night before his lecture dominated everything.  The observation that
this burst was at cosmological distances, gives one of longest sought
answers in Astrophysics of the last decade and is a quantum leap in
our understanding of the enigmatic GRBs. With those observations at
hand we can now for the first time test the quantitative predictions
of models, like the one presented here.

\subsection{TeV \& $\nu$-Astronomy (Rhode, Bean)}

Besides GRBs, Blazars---presumably black hole powered relativistic
jets, pointing at the observer---are the other main source of
high-energy photons. They have now been observed at TeV energies,
where they sometimes have their strongest electromagnetic output and
also demonstrate extreme variability down to small fractions of an hour.

The key point of these TeV observations of Blazars is to decide which
emission mechanism is the dominant one: are hadronic or leptonic
interactions the basis for the high-energy emission?  In the hadronic
interaction model protons are accelerated to high energies.

Here, we assume a shock which accelerates protons and electrons and
wish to consider what the maximum photon energy can be before
Klein-Nishina effects or other losses cut off the spectrum for an
observer.  The shock velocity is taken to be $U_1$, the magnetic field
strength $B$, the fraction of the homogeneous magnetic field energy in
turbulence $b$, the natural logarithm of the wavenumber $k$ range
ratio of turbulence as 
$\Lambda = \ln (k_{max} / k_{min})$, then the
maximum Lorentz factor for protons can be written as \cite{BS87}:

$$
\gamma_{p, max} \; = \; 7.4 \, 10^{10} \, (\frac{b}{\Lambda})^{1/2} \,
\frac{1}{B^{1/2}} \, \frac{U_1}{c}
$$ Here the maximum energy is limited by synchrotron losses and we have
adopted a saturated spectrum of turbulence with $k^{-1}$, different
from Biermann \& Strittmatter \cite{BS87}.  We note, that using the limit of
a relativistic shock velocity and setting all
other parameters to their limits corresponds to setting the
acceleration time scale equal to the Larmor cycle time, which would
seem to be a rather extreme limit; however, in order to derive a
strong upper limit, we will go to this limit below (see also the
arguments on electron maximal energies regarding the Crab nebula by de
Jager and Harding \cite{Jager92}).  Of course, a strongly relativistic
shock would modify this argument as seen from an outside observer
\cite{Peacock81}. Therefore, going to this absolute limit we will set
$
(b/\Lambda)^{1/2} \, (U_1/c) \; = \; 1.
$
The expression for the maximum Lorentz factor of electrons is
$
\gamma_{e, max} \; = \; 1.3 \cdot 10^7 \, B^{-1/2}$.
The Klein Nishina cutoff becomes important when the photon energy
in the frame of the collision approaches the rest mass of the
electron, and so when
$
\gamma_e h \nu \approx m_e c^2.
$

Including the Lorentz-factor of the bulk motion $\gamma_j$ we obtain
$$
(h \nu)_{lept, max} \; \approx \; 6 \cdot 10^{13} \, {\rm eV} \,
B^{-1/2} \, (\gamma_j/10).
$$
Now, the magnetic field strength is expected to approximately scale as
$
B \; \approx \; 10^4 \, {\rm Gauss} \, (r/3 \cdot10^{13} \, {\rm cm}).
$

Therefore, going to this rather extreme limit, and using the region close to
the black hole, the maximum photon energy comes out to be

$$
(h \nu)_{lept, max} \; \approx \; 6 \cdot 10^{11} \, {\rm eV} \,
(\frac{10^4 \, {\rm Gauss}}{B})^{1/2} \, \frac{\gamma_j}{10}.
$$

This is in fact quite close to the maximum derived in various models in
the literature, {\it e.g.},
\cite{DS92,DS93,DS94,Boettcher96,Sikora97}.  Therefore, should the observations
conclusively demonstrate that the spectrum continues straight to above
10 TeV photon energy (for which there may be some indication), then
the leptonic models would begin to have serious difficulties.

Going through the same argument with hadronic processes, we obtain
$
(h \nu)_{max} \; \approx \; m_{\pi} c^2 \gamma_{p, max}
$
and so for the same parameters as above, this maximum is

$$
(h \nu)_{hadr,max} \; \approx \; 7 \cdot 10^{19} \, {\rm eV} \, 
\frac{1}{B^{1/2}} \, \frac{\gamma_j}{10}.
$$

As a result, hadronic processes are much more readily able to account
for photon energies in the high TeV range, {\it e.g.},
\cite{K1,K2,MB89,MKB,KMa,KMb,KMc,KMd,MDSB91,MSB92,NMB,NiBia}.
However, the present observations do not yet allow a final judgment
on this question.

Protheroe \& Biermann \cite{Protheroe96d} argued recently that the infrared
radiation field from the molecular cloud torus, expected to exist in all AGN in
the ``Unified Scheme", will terminate all TeV photons unless they are emitted
above the torus, that is a distance from the central engine of order 0.1 to 1
pc.  This argument is safe by a factor of 1000, {\it i.e.} even if the 
luminosity of the torus is 1000 times weaker, TeV photons still have a 
hard time to escape.

This means that many models in the literature would fail and especially the
leptonic models would have problems.  However, the feature used in many of
these models, a small distance to the central engine, is not always essential,
and so we can expect this class of models to get rejuvenated.

The final judgment will come from the high energy neutrino
observations, which are a firm prediction of the hadronic models.  In
this vein, Wolfgang Rhode reviewed the current situation:

Within the last few years Blazars have been investigated with
different techniques (satellites, Cherenkov telescopes, air shower
arrays) above photon energies of more than 1 GeV. The high energy part
($>$10 TeV) of the spectrum is of special interest, as noted
above. First, the detection of photons far above 10 TeV would favor
accelerated protons as the primary high energy particles. Secondly,
the distance up to which a $>$10 TeV photon can be observed depends on
the density of infrared photons at $\approx \, 10^{-2}$ eV in the
universe.  An observation of a distance dependence of the high
energetic photon flux from Blazars at various distances could thus
provide the possibility to measure directly the cosmological density
of the infrared (IR) background radiation
\cite{Stecker93},  and therefore provide a crucial check on the early evolution
of galaxies and their activities, both in their starburst mode as well as in
their central activity \cite{StanevF97}.

It appears that modeling the observed gamma-ray background is
consistent with a flat photon spectrum of index -2 \cite{Sreekumar97}
for the AGN responsible for the background, consistent with a typical spectrum
for Blazars. 

Mannheim {\it et al.} \cite{Mannheim96} investigated systematically a
sample of 13 Blazars within the field of view of the HEGRA (High Energy
Gamma Ray Astronomy) experiment, which were all close enough
(z $<$ 0.1) to be not absorbed at photon energies of several ten TeV.  This
sample was later enlarged to 30 sources.

The Whipple group discovered two Blazars, Mrk 421 and Mrk 501, at TeV
energies \cite{Punch92,Quinn96}. These observations were confirmed by
the two prototype telescopes (threshold about 1.5 TeV) of the HEGRA
telescope array during the setup phase of the array
\cite{Petry96,Petry97}. Both groups also looked at other sources and
established low upper limits \cite{Rhode97b}. In spring 1997 Mrk 501
was observed to show an eightfold flux enhancement over the last two
months before the meeting (May 1997) which was still
continuing\cite{IAU97}.

The HEGRA telescope system observed Mrk 421 up to more than 5 TeV
\cite{Krennrich97}, and Mrk 501 up to about 20 TeV
\cite{Hermann97,Petry97}.

As reported elsewhere by a team led by Stefan Wagner
\cite{Wagner97a,Wagner97b} a campaign to simultaneously monitor the BL
Lac object Mrk 421 at optical, X-ray and Gamma-ray wavelengths was done
together with the ASCA-satellite and the Whipple observatory.  The
variations are correlated on time scales of one to several days.  In
one case the optical brightness has been observed to vary within 60
seconds, putting extreme constraints on any theoretical model.

Besides the two now famous sources, the sample of 30 Blazar sources was
further investigated by the HEGRA team with several independent data
sets and analysis techniques \cite{Magnussen97}. In autumn 1996 a
possible detection of a cumulative signal and a marginal detection of
the source 0116+319 in three of five data sets were reported by the
HEGRA collaboration \cite{Meyer96,Rhode97a}. A test of the z
dependence of the significance of the detection of all sources of
these data sets suggests, that the detected TeV photons for individual
Blazars with z$>$0.07 was consistent with zero significance, while the
distribution of detection significances for TeV emission from z$<$0.07
Blazars was above zero---even though they individually did not reached
the 5$\sigma$ level.  As already pointed out in \cite{Meyer96} two
other HEGRA data sets do not show this behavior.  Such different
results, nevertheless, are still consistent with the present knowledge
of the highly variable time structure of the sources.

Including all systematic uncertainties, the mean maximum photon energy
of the Blazars as a combined sample is expected to be between 30 TeV
and 70 TeV.  At higher energies a detection of signals from this
object class would require a large surface array
\cite{Mannheim96,Westerhoff95}.

The detection of photons of Mrk 501 up to about 20 TeV now shows that
the calculations of the infrared background given in \cite{Stecker93}
led to a severe overestimate of the IR photon density.  As pointed out
in \cite{Meyer96} an independent upper limit can be tentatively
calculated by using the fact, that only up to z$\simeq $0.07
significant photon detection excesses were found in the Blazar
sample. One obtains a IR-photon density close to the lower estimates
of MacMinn and Primack \cite{MacMinn96}. This conclusion has recently
been confirmed by a calculation of Berezinsky {\it et al.}
\cite{Berezinsky97}.

The high energy TeV photons observed with both detector types
(Cherenkov telescopes and surface array) seem presently to suggest
that hadronic processes are required in Blazars; we note once again,
that high-energy protons are in fact directly observed here on earth.

An ultimate test for many of these models is the detection of high
energy neutrinos. Wolfgang Rhode briefly discussed some ramifications
of the recent discovery \cite{Hunter97} that the canonical model to
explain the gamma-ray emission spectrum of our Galaxy---assumed to be
produced by Cosmic Rays impinging on thermal gas and dust in the
Galaxy---provides an excellent fit to the spatial variations, but a
contradiction with the spectrum.  It appears as if the cosmic ray
spectrum responsible for the interaction (p-p collisions leading to
pions which decay) is quite a bit flatter than the spectrum which is
both indicated by direct observation and by radio observations of
other galaxies. In an extensive collaboration involving groups at
NASA, at Bartol, in Wuppertal and in Bonn, it has been shown, that
using this inference one can predict anew the neutrino spectrum of
the Galaxy giving a flux of neutrinos about ten times higher than
expected so far.  Hence through gamma-ray observations we can learn
directly something about the in-situ properties of high-energy
particles far away from earth.  \vskip0.5cm

As discussed by Francis Halzen in this volume, intense efforts
are under way (AMANDA) to directly detect extragalactic neutrinos from
hadronic processes in the universe. In this workshop Alice L. Bean
reported on an alternative method to detect these by radio emission
from air-showers in ice---the RICE experiment
\cite{Allen97}.

Consider a high energy ($>$100 TeV) neutrino coming from underneath
the South Pole ice sheet, and causing an electromagnetic shower.  This
shower causes a radio emission pulse to travel in a cone-shaped
surface through the ice.  In the RICE experiment radio antennas have
been lowered into the ice, and can pick up any such radio emission.
RICE will consist of an array of compact radio (100 to 1000 MHz)
receivers buried in the ice at the South Pole.  During the 1995-96 and
1996-97 austral summers, several receivers and transmitters were
deployed in bore holes drilled for the AMANDA project, at depths of
141 to 260 m.  At present Alice Bean {\it et al.} are testing the setup with
a receiver and try to develop algorithms to pick up any signal from
the background.  Only in coincidence with the AMANDA array can such a
nice and old idea be fully developed and properly tested.  If it
works, it will be a great boost, because radio technology is well
developed and mature.

\subsection{Cosmic Rays (Biermann)}


The absolute record holder in energy, however, are high-energy cosmic
rays, for which the origin and transport through the intergalactic magnetic
field were discussed by Peter Biermann:

The recent detection of several cosmic ray events with energies beyond
$10^{20}$ eV is challenging astrophysical theories
\cite{Venyabook,Gaisser90,Sigl95,Biermann97a,Nagano97}.
Theoretically, the microwave background does not allow particles
beyond $\approx 5 \cdot 10^{19}$ eV to reach us from cosmological
distances
\cite{Greisen66,ZK66,Stecker68,AC,BG88,UHECRI,Geddes96,Rachen96}.
However, we now have a significant number of clear events of particles
with energies beyond this limit even though these highest energy
cosmic rays clearly cannot be contained in our Galactic disk and
therefore must originate further outside.  Biermann briefly
sketched the various proposals to explain these high energy events,
such as monopoles \cite{Weiler95}, the decay of exotic particles
\cite{BHS,Sigl95,Daum95,Rhode95,Karle95a,Proth96b,Protheroe96c},
shocks in the large scale structure of the universe
\cite{Norman95,Kang95}, compact objects \cite{Hillas84,Shemi95},
Gamma-ray bursts (see above), large scale shocks in our Galactic halo
\cite{Morfill87}, galaxy collisions \cite{CesPt93}, clusters of
galaxies \cite{Kang96,Ensslin97}, active galactic nuclei
\cite{BS87,Proth92}, and, specifically, radio galaxies
\cite{BS87,RS91}.  In any model in which the cosmic rays arrive from
nearby cosmological distances, say, from sources related to galaxies,
we can make some strong predictions: The clustering of arrival
directions on the sky ought to correspond to the source clustering for
energies at which intergalactic scattering by magnetic fields is no
longer important, and for which the cosmologically local structure of
the universe is still inhomogeneous.  Above $4 \cdot 10^{19}$ eV the
arrival directions of cosmic rays, as seen by the Haverah Park array
\cite{PRL95}, the Akeno array \cite{Hayashida96}, and also by a
combination of all experiments \cite{Uchihori97}, are no longer
isotropic, but appear to partially cluster towards the super-galactic
plane, the locus of cosmologically nearby normal galaxies, and radio
galaxies.  Some local enhancements of the very high-energy cosmic rays
may be due to several identifiable radio galaxies; one such candidate
is the radio galaxy 3C134 \cite{3CRRb}.

On the other hand, using the known distribution of candidate sources such as
radio galaxies \cite{UHECRI,CRIV,UHECRII},
and not just the simplified notion of
the super-galactic plane \cite{deV56,deV75a,deV75b,deV76,ShaverP89,Shaver91},
one can simulate the clustering of arrival directions
\cite{Waxman96,Stanev97}:  One
result is that the clustering to the super-galactic plane as derived
from the source locations should be weak, and might not be seriously
detectable with the statistics available.

Clearly the transport of Cosmic Rays \cite{Proth95} through the
inter-galactic medium and knowledge of cosmological magnetic field 
strengths is important. Simulations
\cite{Biermann97b,Biermann97c} of the formation of cosmological
structure, for example allow to determine the spatial inhomogeneity of
cosmic magnetic fields.  Such simulations, however, do not give an
absolute number for the strength of the magnetic field. Combining
these simulations with observations of the ``Rotation Measure'' (of
polarized light) to distant radio sources allows then to deduce upper
limits for the strength of the magnetic field.  These upper limits are
of order $0.2 \, {\rm to} \, 2
\; \mu {\rm gauss}$ along the filaments and sheets of the galaxy
distribution.  In one case, the sheet outside the Coma cluster, there
is a definitive estimate of the strength of the magnetic field
consistent with this range
\cite{Kim89,Kronberg94}.  Such
estimates are almost three orders of magnitude higher than hitherto
assumed, usually based on derivation using the same data, but assuming
a {\it homogeneous} universe (which we know is wrong).  High energy
cosmic ray particles can be either captured or strongly scattered in
such magnetic filaments and sheets, depending on the initial
transverse momentum.  The cosmological background in radio and X-ray
wavelengths will have contributions from these intergalactic filaments
and sheets, should the magnetic fields really be as high as $0.2 \,
{\rm to} \, 2
\; \mu {\rm gauss}$.  We conclude that the  magnetic field structure in the
universe is likely to be just as extremely inhomogeneous as the galaxy
distribution.  Hence, some fraction of highly energetic particles should
be trapped inside the sheets of the baryonic matter distribution, and
so produce the weak correlation with the super-galactic plane detected.

\section{Relativistic Astrophysics}

\subsection{Black Holes and Relativistic Jets (Falcke, Wiita)}

\mbox{Heino~Falcke reported on {\bf Jets in AGN}}:
Astrophysical jets can be the largest and most impressive
signs of the energetic phenomena one commonly associates with black
holes and active galactic nuclei (AGN) and Heino Falcke reviewed our
continuously expanding picture of those gigantic cosmic plasma
accelerators.

In recent years it was found that, rather than only a minority of
sources, almost all types of black holes seem to produce those
outflows. Besides the well known radio-loud quasars and radio galaxies
this includes radio-quiet quasars, Seyferts, LINERs, and X-ray
binaries (stellar mass black holes) as well. Those jets can
substantially influence their environment and are often the site for
intense energetic phenomena, {\it e.g.} the production of gamma-rays and
cosmic-ray particles of the highest energies known today. The
energy-budget of those large, extended jets is probably mainly
controlled by the accretion rate onto the central object and a still
mysterious effect that causes a dichotomy in the jet emissivity
relative to the accretion power. Heino Falcke discussed the evidence
for astrophysical jets in various classes of AGN and their basic
parameters which are crucial for the modeling of all energetic
phenomena that have been linked to AGN.

It is often argued that the escape
speed from the central object is an important factor that determines
the terminal jet speed. If that is true and since we believe that most
of the AGN are powered by a black hole one may expect
that if an AGN produces a jet it should {\it always} be
relativistic. Consequently the crucial question then becomes: Which
classes of AGN have jets? In \cite{Falcke94} and \cite{Falcke95a}
the authors wrote down a hypothesis, simply
stating that since black holes do not have many free parameters, AGN
should be similar in their basic properties (``the universal engine'',
\cite{Falcke96a} and hence one should {\it ab initio} assume that all AGN have
relativistic jets rather than only a few sub-classes. As it turned out, this
hypothesis, in its simplicity, was surprisingly successful.

However, there are interesting difficulties: One finds a clear
dichotomy between radio-loud and radio-quiet quasars and it is often
assumed that the radio-quiet quasars do not have a relativistic jet at
all, while VLA observations of the steep-spectrum radio-loud PG
quasars \cite{Miller93} and \cite{KIK94} have clearly established,
that those sources have large scale radio jets.  Heino Falcke
therefore asked: what would be the consequences, if radio-quiet
quasars too would have relativistic jets? As for radio-loud quasars,
the most prominent sources would be those which are pointing towards
us and are relativistically boosted. In an optically selected sample,
we would expect that, if radio-quiet quasars have relativistic jets,
some of the quasars are accidentally pointing towards us, thus
producing a population of `weak Blazars' with a number of predictable
properties. And indeed, Miller {\it et al.} \cite{Miller93} and Falcke et
al. \cite{Falcke95b,Falcke96e} were able to identify a small sample of
radio-intermediate quasars (RIQ) which met all the requirements for
being relativistically boosted, intrinsically {\it
radio-quiet} quasars \cite{Falcke96f,Falcke97b}, thus strongly suggesting
that in fact {\it all} rather than just 10\% of quasars have
relativistic jets.

Besides radio-quiet quasars, there is another important regime where
one should find relativistic jets. Quasars are powered by black holes
with high accretion rates, but what happens if the accretion rate
decreases? Will those jets die completely? Ho {\it et al.} \cite{Ho95,Ho97}
found that roughly one half of all nearby galaxies show signs of
optical activity in the nucleus, in the form of LINER or Seyfert
spectra, and quite a number also show nuclear radio emission. This
could indicate the presence of a black hole engine
\cite{Heckman80,Falcke96b,Falcke96c,Falcke97a}.

Falcke and collaborators conclude \cite{Falcke97a} that indeed the
radio cores in LINERs are part of the central engine since optical and
radio fluxes are correlated. Moreover, we can compare the radio and
emission-line luminosities with the jet/disk model by Falcke \&
Biermann \cite{Falcke96d}.  The model predicted a specific
radio/nuclear luminosity correlation for low-power AGN and is based on
the assumption that accretion disk luminosity and jet power in AGN are
coupled by a universal constant. The LINERs fall exactly into the
range predicted for low-luminosity, {\it radio-loud} jets.

This result not only strongly suggests that LINERs do have powerful
nuclear radio jets
({\it e.g.} M87; NGC4258, \cite{Herrnstein97}; M81, \cite{Bietenholz96}, etc.)
but is also consistent with mildly relativistic Lorentz factors around
$\gamma_{\rm j}\simeq2$ as used in the model. That should be compared with
Lorentz factors of $\gamma\simeq6-10$ derived with the same method for
radio-loud quasars \cite{Falcke95b}.

\vskip0.5cm

Paul Wiita reported on work done with Gopal-Krishna and Vasant
Kulkarni (both at the NCRA of the Tata institute at Pune, India) on
the status of the so called ``Unified Scheme" which is used to explain
the differences of the most powerful radio jets just in terms of
different orientation and obscuration:

A key argument in favor of orientation based unification schemes is
the finding that among the most powerful 3CRR radio sources the
(apparent) median linear size of quasars is smaller than that of radio
galaxies, which supports the idea that quasars are a subset of radio
galaxies, distinguished by being viewed at smaller angles to the line
of sight.  Recent measurements of radio sizes for a few other low
frequency samples are, however, not in accord with this trend, leading
to the claim that orientation may not be the main difference between
radio galaxies and quasars.  Wiita pointed out that this
``inconsistency'' can be removed by making allowance for the temporal
evolution of sources in both size and luminosity, as inferred from
independent observations.  This approach can also readily explain the
other claimed ``major discrepancy'' with the unified scheme, namely,
the difference between the radio luminosity--size correlations for
quasars and radio galaxies.  Some of this work is reported in
\cite{GK96a,GK96b,GK96c}.

\vskip0.5cm

{\bf Black Holes:} The scales of relativistic jets are still large
compared to the event horizon of a black hole, but even black holes
are slowly coming into sight! {\it E.g.} Falcke pointed out that the mass of
the black hole in the center of our Galaxy is now determined with a
very a high precision ($2.65(\pm0.2)\cdot10^6M_\odot$) by the
measurement of proper motion of stars \cite{Eckart97}, pushing the
central dark mass density to $10^{12} \, M_\odot$pc$^{-3}$ ({\it i.e.}  the
mass of a whole galaxy concentrated in less then the volume between
the solar system and the next stars).  Any alternatives to a black
hole, {\it e.g.} an ultra compact cluster of stellar remnants seems to be
ruled out now \cite{Maoz97} thus making the Galactic Center source Sgr
A* the best supermassive black hole candidate today. Interestingly,
the mass is so large that the photon horizon of this black hole has a
diameter corresponding to an angular resolution of 27
$\mu$arcsecond. Such a resolution will in fact be reached by planned
VLBI (very long baseline interferometry) experiments operating at 220
GHz.  Since this black hole also has an ultra compact emission region
estimated to be of a similar size, radiating at just this frequency,
this leaves the tantalizing possibility to directly image the horizon
of a black hole at least in one case.

\vskip0.5cm

Paul Wiita also reported on work done with Gang Bao and Ying Xiong,
(also at Georgia State University) and with
Petr Hadrava (at the Astronomical Institute of the Czech Academy)
on Polarization Variability as a Signature of Black Holes:

In regions where electron scattering dominates the opacity
above accretion disks, X-ray radiation originating there
should be partially linearly polarized.  Both observations
of rapid X-ray variability and theoretical studies suggest
that this inner disk region is unstable and could appear
clumpy. He showed how variations in the orbital parameters of the
bright spots and the angle between the line of sight and
the disk axis affect the observed polarization.
The amplitudes of both the changes in the degree of polarization and
the angle of the plane of polarization are energy-dependent. They
are relatively independent of the physical mechanism producing the
polarization. This feature is directly created by the gravitational
bending of light rays by the central black hole and it is apparently
unique to a system including a black hole and an accretion disk.
This work is reported in \cite{Bao97a,Bao97b}.

\vskip0.5cm

\subsection{Ultrahigh magnetic fields and extreme densities (Kennedy)}

Dallas Kennedy reported on work done with K. S. Gopinath (U. Florida) and
J. M. Gelb (U. Texas) on relativistic Landau states of electrons in
intense astrophysical magnetic fields, encountered especially in neutron
stars at $\gsim$ $10^{12}$ Gauss (LANL astro-ph/9702014 and astro-ph/
9703108).

The classical and semi-classical orbits of relativistic charged particles
were outlined for motion on a spherical surface, in an intense magnetic
dipole background.  The dipole and rotational axes in general should not
be aligned, if the star's magnetic dynamo is to be self-sustaining.  
Kinematic regimes differ depending on the relative sizes of energy,
canonical azimuthal angular momentum, and magnetic field strength 
in rescaled units.  Magnetic flux enclosed by the orbits is quantized very
close to the poles.  Open questions relating to the state of electronic
matter near neutron star surfaces were sketched.

Subsequent work has extended this calculation to the full 3-D problem at
finite density, with electrons in local field-transverse planes, but
including the gravitational and hadronic structure only as given
backgrounds (LANL astro-ph/9707196).

Further questions include the magnetically-induced structural changes below 
the star surface and changed state of matter, as well as observational
signals of such exotic ``quantum Hall-like'' surface physics.

\section{Cosmology}
A larger and perhaps more entertaining review was given at this
conference by Rocky Kolb. Here we summarize some of
the interesting twists to cosmological models presented during the
astrophysics workshop.

\subsection{Fluctuations in the early universe (Hochberg, Berera)}

David Hochberg reported on
Course-Graining, Structure Formation and the Transition to
Large-Scale Homogeneity in the Universe
\cite{Gurbatov85,Gurbatov89,Berera94,Goldman96,Hochberg96,Barbero97}:

Newtonian hydrodynamics plus FRW cosmology combine to yield a good
description of the matter dominated Universe at large scales. Under
assumptions of vorticity-free flow and validity of the Zeldovich
approximation (that the gravitational acceleration is parallel to the
velocity) the ensuing dynamics can be re-cast in terms of a
cosmological version of the (massive) Kardar-Parisi-Zhang (KPZ)
equation, which has enjoyed extensive application in the study of
surface growth phenomena. Here, he applied it to the problem of the
growth and distribution of large-scale structure in the
Universe. Using the well established techniques of the dynamic
re-normalization group to study the scaling properties of the solutions
of the KPZ equation, he calculated the power law behavior and
attendant exponent of the galaxy to galaxy correlation function and
showed that the transition to large scale homogeneity is an necessary
consequence of the course-graining.

\vskip0.5cm

Arjun Berera reported on an attempt to determine
the largest scale of primordial density perturbations
beyond the Hubble radius with COBE-DMR and the implications for early
universe cosmology:

Causality imposes rigorous constraints which suppress
super-horizon scale coherence. The power spectrum of scalar primordial
density perturbations is modified by inclusion of a super-Hubble
suppression scale in order to respect this constraint mandated by
causality.  A recent analysis of COBE-DMR data was presented, in which
measurements were made of the super-Hubble suppression scale, the
spectral index and the amplitude.  Theoretical implications of this
analysis focus on the warm inflation scenario, which in part motivated
this COBE analysis.  A summary of the scenario was presented which
included discussion about self-consistency, avoidance of a re-heating
period, relevance to open universe and primordial seeds of density
perturbations and magnetic fields.

Although the details of the scenario are model dependent, this so
identified regime is an outcome only of Friedmann cosmology. He asked
if this modification would have any significant effects on the
mechanisms to produce large magnetic fields.

\vskip0.5cm

As mention in the previous Cosmic ray section, the origin of magnetic
fields is still an elusive goal in our understanding of the universe;
many models have been proposed, from dynamos working in stars, in
accretion disks, in entire galaxies, and in large scale accretion
flows; primordial magnetic fields have many attractive features, since
they may obviate some of the difficulties faced by the dynamo models.

\subsection{Magnetic Fields in the Universe (Kronberg)}

Recently improved instrumental capabilities over the past decade or so
have greatly improved our knowledge of the extent and strength of
cosmic magnetic fields. Some surprising discoveries have emerged and
Phil Kronberg summarized our latest knowledge of the strength and
extent of magnetic fields in galaxies, galaxy clusters, and what
little inkling has been gained about widespread intergalactic magnetic
fields.  The status of our knowledge is well described by Kronberg
\cite{Kronberg94,Kronberg97} in two excellent reviews.

Phil Kronberg included some basic theory of magnetic field
regeneration, and some constraints imposed by recent experimental
data. Kronberg also explained the observational methods of probing
magnetic fields in outer space, and mentioned future prospects linked
to experiments with gamma- and cosmic ray detectors, and at low
frequency radio and sub-millimeter bands.

The information came from measurements of the Faraday rotation of the plane
of linear polarization turned by the transfer through an ionized and
magnetized medium.  This rate of angle rotation with increasing wavelength
is defined as Rotation Measure (RM) (radians/m$^2$), and is proportional to the
integral  of the electron density along the line of sight, folded with the
parallel magnetic  field components.  On the sky most of the effects are
dominated  by our own Galaxy.   Looking further one finds the following:

a) In normal galaxies such as our own the magnetic fields are of order
3 to 10 $\mu$gauss \cite{Beck96}, with a lower limit of about 1 $\mu$gauss.

b) In the halo of our Galaxy the magnetic field is $\approx$ 0.2 $\mu$gauss,
over a scale of a few kpc \cite{Han94}.

c)  In starburst galaxies such as M82 the magnetic field can be very much
stronger, with values up to $\approx$ 30 $\mu$gauss possible \cite{KBS85}.

d) In clusters of galaxies the magnetic field strength is subject to
some uncertainty in the number of magnetic field reversals along
lines of sight through the cluster
\cite{Kim90,Kim91,Kronberg94}, but the overall rotation of the polarization
plane of background radio sources suggests values of a few $\mu$gauss, while
some  cooling-flow clusters with strong radio galaxies are known to have
values about ten  times higher.

e) Outside one cluster, the Coma cluster, Kim {\it et al.} \cite{Kim89}
have estimated the strength of the magnetic field near 0.1 $\mu$gauss,
assuming  equipartition between magnetic fields and relativistic particles.

f) Across cosmological distances there is only an {\it upper limit}
of $\approx$ 1 nanogauss, derived from Rotation Measure data measured along
lines of  sight to cosmologically distant radio-quasars \cite{Kronberg94},
assuming a reversal scale of the magnetic field structure of 1 Mpc, and
otherwise  cosmologically homogeneous properties of the intergalactic
medium.  If we were to  assume that the reversal  scale is the  same as the
bubble structure of the galaxy  distribution, then
this limit would be $\approx$ 200 picogauss.

g) We do not (yet) have a successful theory to account for the origin
of cosmological magnetic fields, neither in galaxies, nor in clusters.
Stars such  as the Sun show evidence  for a fast dynamo acting to reverse the
magnetic field every 11 years, and  so it is often assumed, but has not finally
been  demonstrated, that galaxies do the same on the much larger scale
\cite{Beck96}.   In clusters of galaxies, the magnetic field can energetically
be provided by the radio galaxies \cite{Kronberg94,Ensslin97}.

The observation of normal magnetic fields in a galaxy at fairly large
redshift suggests that magnetic fields can build up over
cosmologically fairly short time scales \cite{Kronberg92}.  On the
other hand, the strength and topology of magnetic fields in galaxies
appears to exclude an origin as simply primordial
\cite{Parker92,Beck96}.  One recent attempt to simulate the growth of
cosmological magnetic fields has been made by Kulsrud and
collaborators \cite{Kulsrud96}, using the structure formation itself
as the source in a battery process \cite{Biermann50}.  Another
argument has been that massive stars and compact accretion disks in
AGN may be all that is needed, also starting with a battery process
\cite{Biermann97a}.

\subsection{Cosmological Birefringence (Ralston)}

Finally, Ralston reported on his much debated claim of cosmological
birefringence, published with Nodlund.

Ralston asked the question, whether there is bi-refringence of the
universe, based upon an analysis of radio galaxy data \cite{Nodlund97a}. If
so, this would have been a major change for our understanding of the
universe \cite{Kuehne97}.  He finds a big effect for redshifts larger than 0.3
from a large database of previously published measurements of polarization
vectors of galaxies.  While the proposal has spawned an intense debate
with a number of counter examples on the LANL electronic preprint server
\cite{Wardle97,Carroll97,Eisenstein97,Nodlund97b,Ralston97}
and in the popular press, Ralston himself concluded ``Barring hidden
systematic effects, the analysis indicates a new cosmological effect".
Clearly the debate demonstrates, that most researchers still would put
more weight on the first part of the sentence and favor a systematic
effect in the analysis of the data to interpret Nodland's \& Ralston's
findings, before making drastic changes to existing cosmological
models.

\section{SUMMARY}

Compact objects from the centers of the Gamma Ray Bursts to the the
presumed black holes at the focus of the AGN span about ten powers of
ten in mass.  Their associated outflows, whether explosive,
non-steady, steady or quiescent, have been the focus of much work over
the past several decades.  Those are the laboratories where new
physics can and will be learned. It appears that basically the same
concepts may help us to understand much of what we observe: Everywhere
we look at a compact object such as a black hole, there seems to be an
accretion flow (disk) and a jet, with gigantic and relativistic shock
waves running through these jets, and accelerating particles.  Another
common analogy is simply the formation of a compact object with a
concurrent explosion, be it as a supernova, or as a Gamma Ray Burst.
In the case of a Gamma Ray Burst again a relativistic flow seems
indicated, and we come back to the basic AGN language, and maybe the
same physics.

Many questions still remain: Are the ZeV energy particles, presumably
protons, really derived from shock waves in such jets from the most
powerful radio galaxies we know?  Or do they come from the Gamma Ray
Bursts?  Or, most excitingly perhaps, do they come from the decay of
particles of GUT-scale energies?

Are the emissions that we observe in Gamma Ray Bursts or in TeV Blazars, all
derivable from leptonic processes, or do they require hadronic processes to
get started? What will neutrino astronomy bring us?

Can we connect all these observations to the structure formation of the
universe, whether it is the cosmic magnetic fields or the first seeds of black
holes?

These and many more tasks are waiting for us and thanks to impressive
developments in recent years, {\it we can be certain that  the ``Golden
Years of Astrophysics'' are not over yet}.

\section*{ACKNOWLEDGMENTS}

PB and HF wish to thank all participants for the discussions at the
meeting as well as before and afterwards.  For this article we have in
part used some of the material the participants have generously
provided to us. PB also wishes to thank Karl Mannheim, Giovanna Pugliese and
J{\"o}rg Rachen for comments, as well as Abhas Mitra, Tsvi Piran,
Ray Protheroe, G{\"u}nther Sigl, and Todor Stanev 
for intense discussions on various topics also
covered in this workshop.


\end{document}